\documentclass{aastex62}

\graphicspath{{./}{figures/}}

\submitjournal{ApJ}

\shorttitle{Dust echo tomography with bright Galactic X-ray bursts}
\shortauthors{Corrales et al.}

\newcommand{\NH}{{\rm N}_{\rm H}}
\newcommand{\SGRecho}{1E~1547.0-5408}
\newcommand{\VCygni}{V404~Cygni}
\newcommand{\Cir}{Cir~X-1}
\newcommand{\XMM}{{\sl XMM-Newton}}
\newcommand{\Swift}{{\sl Swift}}
\newcommand{\Chandra}{{\sl Chandra}}

\newcommand{\fcgs}{\mathcal{F}_{\rm cgs}}

\begin{document}

\title{The X-ray variable sky as seen by MAXI: the future of dust echo tomography with bright Galactic X-ray bursts}

\correspondingauthor{\email{liac@umich.edu}}

\author[0000-0002-5466-3817]{Lia Corrales}
\affil{LSA Collegiate Fellow, University of Michigan, 
Ann Arbor, MI 48109, USA}
\affil{Einstein Postdoctoral Fellow}
\affil{University of Wisconsin - Madison, 
Madison, WI 53706, USA}

\author{Brianna S. Mills}
\affiliation{University of Wisconsin - Madison, 
Madison, WI 53706, USA}
\affil{University of Louisville, 
Louisville, KY 40292, USA}
\affil{University of Virgnia, 
Charlottesville, VA 22904, USA}

\author{Sebastian Heinz}
\affiliation{University of Wisconsin - Madison, 
Madison, WI 53706, USA}

\author{Gerard M.~Williger}
\affil{University of Louisville, 
Louisville, KY 40292, USA}
\affil{Jeremiah Horrocks Institute, University of Central Lancashire, Preston, PR1 2HE, England}
\affil{Institute for Astrophysics and Computational Sciences, Catholic University of America, Washington, DC 20064}
\affil{Konkoly Observatory, 15-17 Konkoly-Thege Miklós út, 1121 Budapest, Hungary}

\begin{abstract}

Bright, short duration X-ray flares from accreting compact objects produce thin, dust scattering rings that  enable dust echo tomography: high precision distance measurements and mapping of the line-of-sight distribution of dust. 
This work looks to the past activity of X-ray transient outbursts  
in order to predict the number of sight lines available for dust echo tomography.
We search for and measure the properties of 3$\sigma$ significant flares in the 2-4~keV light curves of all objects available in the public MAXI archive. We derive a fluence sensitivity limit of $10^{-3}$~erg~cm$^{-2}$ for the techniques used to analyze the light curves. This limits the study mainly to flares from Galactic X-ray sources. 
We obtain the number density of flares and estimate the total fluence of the corresponding dust echoes. However, the sharpness of a dust echo ring depends on the duration of a flare relative to quiescence. We select flares that are shorter than their corresponding quiescent period to calculate a number density distribution for dust echo rings as a function of fluence. The results are fit with a power law of slope $-2.3 \pm 0.1$.
Extrapolating this to dimmer flares, we estimate that the next generation of X-ray telescopes will be 30 times more sensitive than current observatories, resulting in 10-30 dust ring echoes per year. The new telescopes will also be 10-100 times more sensitive than {\sl Chandra} to dust ring echoes from the intergalactic medium.

\end{abstract}


\section{Introduction} \label{sec:intro}

Interstellar dust scatters X-ray light over arcminute-scale angles, producing a diffuse scattering halo with an integrated flux $F_{\rm halo}(E) = F_{a} (1 - e^{-\tau})$, where the optical depth to X-ray scattering, $\tau \approx 0.5~E_{\rm keV}^{-2}~(\NH/10^{22}~{\rm cm}^{-2})$, and $F_{a}$ is the absorbed flux of the central X-ray source \citep{PS1995, Corrales2016}.  
Because the scattered light takes a longer path to reach the observer, the observed scattering halo surface brightness profile is a convolution of the dust's line-of-sight position, grain size distribution, and $F_a(t)$ light curve \citep[][and references therein]{Heinz2015}.
When $F_{a}(t)$ takes the form of a single burst with high amplitude and short duration, a scattering halo will appear as a set of discrete rings, where each ring corresponds to a different foreground dust cloud. These rings expand with a characteristic $t^{1/2}$ time dependency that allows X-ray astronomers to map the line-of-sight distribution of dust (``dust echo tomography'') to much higher resolution than currently available with any other method \citep{TS1973,Heinz2015,Heinz2016}. 

Mapping the ISM through dust echo tomography is also important for interpreting the time and spectral evolution of accreting compact objects.
Dust echoes are known to affect the spectral evolution of X-ray variable objects, producing a prolonged soft-tail \citep[e.g.,][]{Pintore2017,Jin2018}. This confusion is particularly acute for X-ray timing missions with low imaging resolution: RXTE, MAXI, and NICER.

To date, the brightest dust echo rings observed have come from four Galactic X-ray sources -- \SGRecho\ \citep{Tiengo2010,Olausen2011,Pintore2017}, \Cir\ \citep{Heinz2015}, \VCygni\ \citep{Vasilopoulos2016,Heinz2016,Beardmore2016}, and 4U~1630-47 \citep{Kalemci2018}. Dust echoes can also be produced by the X-ray components of gamma-ray bursts (GRBs), which scatter off the nearby Galactic medium \citep{Vaughan2004, Vaughan2006, Tiengo2006, Vianello2007, Pintore2017b}. Table~\ref{tab:flares} lists the approximate soft X-ray fluence and ISM column for dust echo rings observed around GRBs and XRBs. 
In most of the GRB cases, fluences were measured from the properties of the dust scattering echo, and the results depend on the adopted grain size distribution.

\begin{table}
\centering
\caption{Flare properties of objects with observed dust echoes}
\label{tab:flares}
\begin{tabular}{lcccl}
\hline
{\bf Object} & {\bf Telescope(s)} & {\bf Fluence}$^a$ & {\bf Gal.}~$\NH$ & {\bf References} \\
	& 	&  (erg~cm$^{-2}$) 	& ($10^{22}$~cm$^{-2}$) & \\
\hline
GRB 031203 & \XMM & $0.7-3 \times 10^{-6}$ & 0.6 & \citet{Vaughan2004,Watson2006} \\
& & & & \citet{Tiengo2006} \\
GRB 050713A & \XMM & $5 \times 10^{-7}$ & 0.1 & \citet{Tiengo2006} \\
GRB 050724 & \Swift & $2 \times 10^{-7}$ & 0.6 & \citet{Vaughan2006} \\
GRB 061019 & \Swift & $5 \times 10^{-7}$ & 0.9 & \citet{Vianello2007} \\
GRB 070129 & \Swift & $7 \times 10^{-7}$ & 0.1 & \citet{Godet2007, Vianello2007} \\
GRB 160623A & \XMM & $2 \times 10^{-6}$ & 0.7 & \citet{Pintore2017b} \\
\hline
1E 1547.0-5408 & \Swift, \XMM & $2-6 \times 10^{-3}$ & 3.0 & \citet{Tiengo2010, Halpern2008} \\
Cir X-1 & \Chandra & 0.025 & 2.0 & \citet{Heinz2015} \\
V404 Cygni & \XMM, \Chandra & 0.01 & 0.6 & \citet{Heinz2016} \\
4U~1630-47 & \Swift, \Chandra & 0.015 & 13.0 & \citet{Kalemci2018}, estimated from MAXI \\
\hline
\multicolumn{5}{l}{$^a$2-4~keV, absorbed fluences estimated from the references cited} \\
\hline
\end{tabular}
\end{table}

One can note from Table~\ref{tab:flares} that the fluences of GRBs producing dust echoes are particularly low. 
In these cases, $\sim 10\%$ of the X-ray light from the flare is deposited into the dust scattering ring. This level rivals the amount of light in the telescopes' point spread function (PSF) wings. However, due to the quick dimming typical of X-ray afterglows, the time delay between the prompt X-ray emission and the dimming afterglow allows the the dust scattering rings to stand out in contrast, even when a central X-ray point source is visible.
In theory, X-ray variability from any high-redshift object can produce echoes that propagate off dust from foreground galaxies or the intergalactic medium, but it requires more sensitive telescopes than currently available \citep{Miralda-Escude1999,Corrales2012,Corrales2015a}. 

Many Galactic XRBs are persistent sources of X-rays, producing a quiescent dust scattering halo. A source that experiences frequent outbursts, or high variation, will create a time variable scattering halo with no clearly defined rings. This study focuses on identifying single, large amplitude outbursts capable of producing thin, high contrast dust echo rings. 

This work evaluates the X-ray light curves from all sources monitored regularly by MAXI \citep[Monitor of All-sky X-ray Image,][]{MAXI2009}, in order to gather the rate of X-ray flares propagating through the interstellar medium (ISM). In Section~\ref{sec:analysis}, we describe the algorithm used to identify flares and discuss its limitations. In Section~\ref{sec:results}, we calculate the number distribution of flares identified in MAXI. A metric for evaluating the likelihood of an outburst to produce sharp ring echoes is discussed in Section~\ref{sec:rings}.  In Section~\ref{sec:future}, we fit a power law to the number distribution of flares and use it to estimate how many X-ray dust ring echoes will be 
seen with the next generation of X-ray telescopes. We also update the results of \citet{Corrales2015a} to estimate the number of X-ray scattering echoes that might be found arising from dust in the intergalactic medium. 
All findings are summarized in Section~\ref{sec:conclusions}.

\section{Data Analysis} \label{sec:analysis}

MAXI 
is the longest operating current all-sky monitor for the soft X-ray band, which is sensitive to dust scattering. The Soft X-ray Large Angle Camera on MAXI captures almost the entire sky on a cadence of 92 minutes, over the 0.7 - 12~keV energy band, with a binned one-day sensitivity limit of 4.5~mCrab ($10^{-10}$~erg~s~cm$^{-2}$) \citep{MAXI2009, MAXI2014}. Because the ISM preferentially removes soft X-rays, the spectral energy distribution of X-ray scattering halos tend to peak around 1-3~keV. We use the publicly available one-day binned 2-4~keV light curves from 398 point sources currently available on the MAXI website\footnote{http://maxi.riken.jp/top/lc.html}, from the start of MAXI operations to MJD 58408, to estimate the probability distribution of soft X-ray flares across the sky.
For the purposes of this work, a flare is defined as any duration longer than several days for which an object's flux is $> 3\sigma$ above its dimmest state.

It should be noted that MAXI is insensitive to most flares that are significantly shorter than a single ISS orbit. Magnetars, Type I neutron star bursts, and X-ray afterglows to GRBs fall into this category. 
As will be demonstrated below, such flares are missed from this study due to limitations of MAXI, not due to data analysis choices.

\subsection{Detection Algorithm}
\label{sec:algorithm}

First, we removed all data points where a monitored object was within 10 degrees of the Sun, which is a large source of contamination. All light curves were smoothed using a 3-day Gaussian convolution, to improve the stability of the  algorithm.  
Limitations imposed by smoothing are discussed in \S\ref{sec:limitations}.

For a baseline flux, we selected the 16th-percentile value from the distribution of flux values within the whole light curve. 
We then subtracted the baseline value from the light curve and calculated the signal-to-noise ratio for each bin. 
Flares were identified by flagging light curve intervals with a signal-to-noise ratio greater than three. 
Flare intervals shorter than five days were discarded. If two flares were separated by an interval shorter than five days, we combined them  into one time interval. This process was repeated so that there was no quiescent period shorter than five days.  
These choices were motivated by analysis of a simulated dataset, described in \S\ref{sec:simulations}. The five day cut-off significantly reduced the number of falsely identified flares.

Finally, we calibrated each light curve by normalizing them with the MAXI light curve for the Crab pulsar, which has a 2-4~keV band flux of $1.1 \times 10^{-8}$~erg~cm~$^{-2}$~s$^{-1}$. We used linear interpolation over data gaps to arrive at a total fluence for each flare interval.

\begin{figure}
\centering
\includegraphics[scale=0.47]{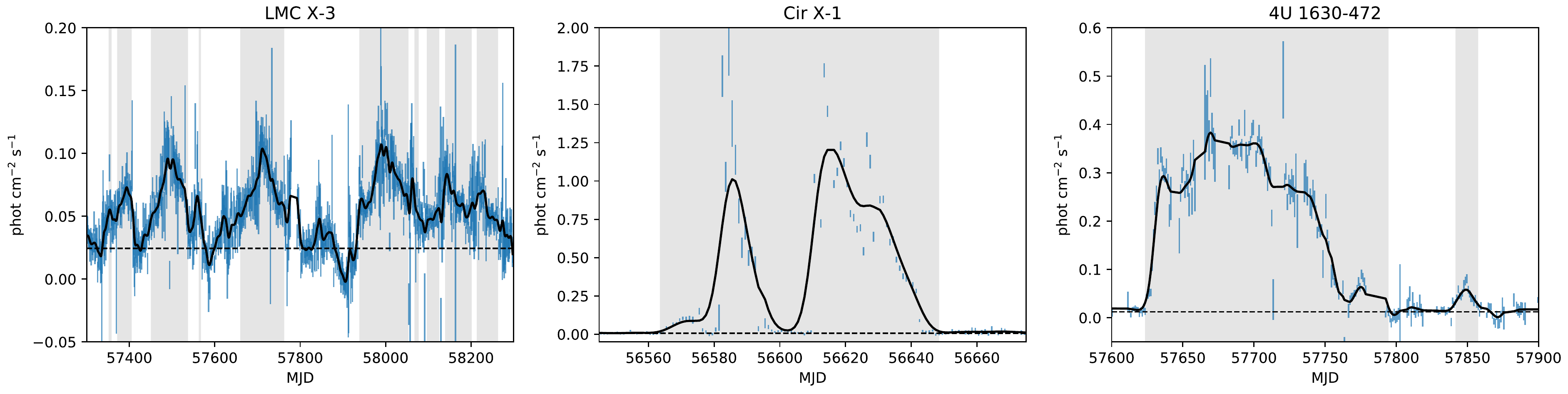}
\caption{
Flare intervals identified from three example light curves. The raw 2-4~keV light curves {\bf supplied by the MAXI website} (blue) are smoothed with a 3-day Gaussian kernel (black). After subtracting a baseline flux value, identified using the lower 1-$\sigma$ value of the dataset (dashed black line), intervals with signal-to-noise greater than three are flagged as flares (shaded grey regions). For X-ray binaries that rarely stay quiescent, such as LMC~X-3 (left), any bright state is flagged as a flare. Two dust-echo producing flares from Cir~X-1 (middle) and 4U~1630-47 (right) are highlighted. The calculated fluences are consistent with {\bf those reported} in the literature.
\label{fig:demo}
}
\end{figure}

Figure~\ref{fig:demo} shows the results for three light curves of interest. LMC~X-3 exhibits erratic behavior with no clearly defined quiescent state. The algorithm flags the intervals when LMC~X-3 is in a bright state. The next two panels show Cir~X-1 and 4U~1630-47 during the flares leading to dust echoes study by \citet{Heinz2015} and \citet{Kalemci2018}, respectively. The calculated fluences are 0.023~erg~cm$^{-2}$ over 85 days (Cir~X-1) and 0.021~erg~cm$^{-2}$ over 171 days (4U~1630-47). These values are consistent with those in the published works. 

We were unable to check on the dust echo producing flares from 1E~1547.0-5408, which occurred before the launch of MAXI, and V404~Cygni, which was only observable by the degraded GSC3 instrument at the time of the flare \citep{Negoro2015}. The publicly available MAXI light curves do not include data from GSC3. 
The MAXI view of V404~Cygni is also affected by source confusion with Cyg~X-1, which is usually much brighter. As a result, the light curve is poorly calibrated and no flares were measured from V404~Cygni.\footnote{Communication with MAXI calibration staff}

\subsection{Sensitivity limits}
\label{sec:simulations}

To examine the accuracy of our analysis, we simulated $1000$ MAXI light curves 
of 900~days long, 
and injected one Gaussian flare into each. 
The reported MAXI sensitivity is 4.5~mCrab for one-day binned data \citep{MAXI2009}, yielding a fluence theoretical lower limit of $5 \times 10^{-5}$~erg~cm$^{-2}$ for a five day long flare.
As such, the flare properties were drawn from a uniform distribution of fluences, $\log(\fcgs) \in [-5, -1]$ (where $\fcgs$ is fluence in units of erg~cm$^{-2}$), and a uniform distribution of Gaussian widths $\sigma ({\rm days}) \in [1, 50]$.\footnote{A 900 day light curve was deemed sufficient to capture flares that are effectively 300~days long (Gaussian $\sigma = 50$~days). As shown later in Section~\ref{sec:rings}, these very long outbursts are typically beyond the scope of interest for dust echo tomography.}
The baseline flux and error bars for each simulated light curve were drawn randomly from three MAXI light curves of objects with a quiescent flux below the sensitivity limits, i.e., those exhibiting a light curve consistent with zero flux throughout: 1ES~1101-23.2, WW~Cet, and VY~Ari. We take these objects as representative of the zero values and error bars arising from the MAXI calibration processes.

\begin{figure}
\centering
\includegraphics[scale=0.75]{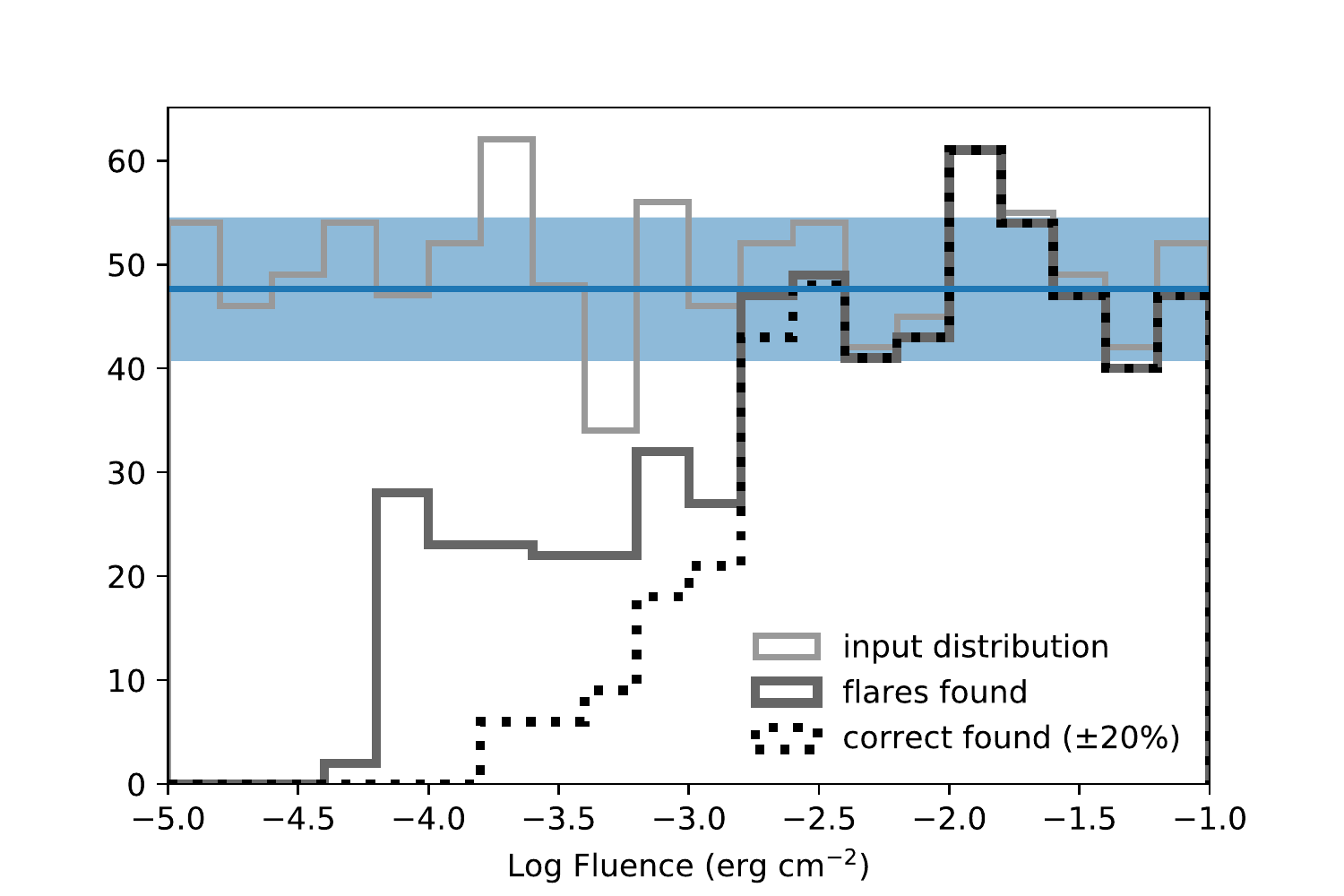}
\caption{A histogram of simulated input (light grey) and the histogram derived with the techniques described in Section~\ref{sec:analysis} (dark grey). The input flares were drawn from a uniform distribution, and the expected $1\sigma$ variation is shaded in blue. 
The dotted black histogram shows how many of the output flares were correct identifications, to within 20\% of the input fluence. 
\label{fig:SimulatedFlares}
}
\end{figure}

Figure~\ref{fig:SimulatedFlares} shows the distribution of detected flares compared to the input distribution. In general, the algorithm returns a large number of short $10^{-4}$~erg~cm$^{-2}$ flares ($<$ 3-5 days) that appear to arise from the noise typical in the MAXI dataset. We were able to cut down on the number of false positives significantly by ignoring flares shorter than five days, and by combining flares that were separated by less than five days. 
The dotted line in Figure~\ref{fig:SimulatedFlares} shows how many flares for which the fluence was correctly retrieved to within 20\%. 
For the subset of flares with fluence $> 10^{-3}$~erg~cm$^{-2}$, our algorithm was able to identify 90\% of all the simulated flares and 100\% of those that were of duration $\leq 20$~days.
We therefore take $\log \fcgs > -3$ as the completeness limit for this study.

Identification of long, high fluence flares is limited by the flux in each bin. We estimated the flux sensitivity of our algorithm by dividing each fluence value by its corresponding duration, yielding 4~mCrab, which is consistent with the expectations for one-day binned MAXI data.

\subsection{Limitations for short outbursts}
\label{sec:limitations}
Using daily binned light curves and three-day smoothing imposes selection effects against short flares. 
In the extreme case of a flare restricted to a single ISS/MAXI orbit (90 minutes), the signal-to-noise of the flare in one-day binned data will be reduced by a factor of $\sqrt{15}\sim 4$ compared to the single-orbit light curve. 
This would eliminate single-orbit flares of signal-to-noise lower than 12 ($< 80$~mCrab) from detection, corresponding to a fluence $< 10^{-5}$~erg~cm$^{-2}$. 
This value is several orders of magnitude below the fluence values of interest for dust echo tomography with Galactic X-ray sources (lower portion of Table~\ref{tab:flares}), which is the main target of this study.

In conclusion, smoothing data has the advantage of reducing the number of false positives, because the variance of the data is significantly reduced.
The critical metric for the detectability of dust scattering echoes of flares is their fluence, which
is preserved in binning and smoothing. 
Section~\ref{sec:simulations} demonstrates that the algorithm retrieved all of the short duration flares with $\log \fcgs > -3$, which is an order of magnitude below our threshold of interest demonstrated by the lower portion of Table~\ref{tab:flares}. 
Thus, the benefits of using binned and smoothed data outweigh the reduction of sensitivity to short flares.

\section{Results and Discussion} \label{sec:results}

Of the 398 sources analyzed, 213 exhibited outbursts that were picked up by the flare detection algorithm. 
To account for source confusion, we evaluated the light curves of three sources within $2^{\circ}$ of each other: SMC~X-1, SMC~X-3, and MAXI~J0057-720. One flare from SMC~X-3 appeared coincidentally in the light curve of MAXI~J0057-720, which is 0.6$^\circ$ away. However, variations from SMC~X-1, which is the brightest of the three and 2$^\circ$ away from the other two objects, did not affect the light curves of either. Thus we chose $1^\circ$ as the threshold for evaluating the effects of source confusion. We identified pairs of sources in the MAXI dataset separated by $< 1^\circ$. Within this subset of light curves, we  searched for flares appearing within 30~days of each other. When coincident flares were found, we kept the larger fluence event and discarded the other. We also visually evaluated the light curves of sources within 2$^\circ$ of the Galactic Center, which hosts a large number of variable compact objects that cannot be resolved with MAXI. The overall process resulted in the removal of 8 flares that were double counted, leaving a total of 854 distinct outbursts with $\log \fcgs > -3$.

\begin{figure*}
\centering
\includegraphics[scale=0.5]{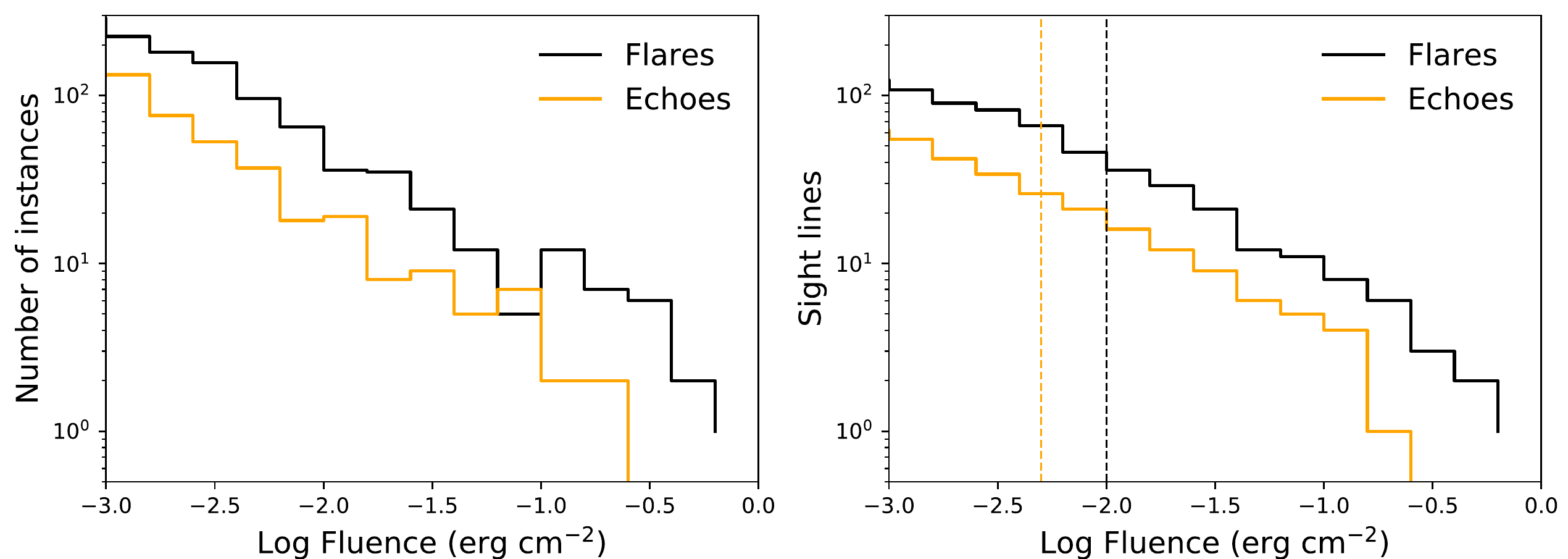}
\caption{
{\sl Left:} A histogram of flares found in 213 MAXI light curves shows the fluence of the flares themselves (black) and the estimated fluence of the corresponding dust echo (orange). 
{\sl Right:} The number of sight lines that exhibit a flare (black) or dust echo (orange) larger than a particular fluence, described on the horizontal axis. This represents the number of sight lines available for performing dust echo tomography. The vertical dashed lines mark the threshold for flares (black) and corresponding echo brightness (orange) above which dust echoes from Galactic XRBs have been observed today. 34 and 24 of the sight lines satisfy two thresholds, respectively. 
\label{fig:FlareHist}
}
\end{figure*}

Figure~\ref{fig:FlareHist} shows a histogram of total number of flares detected as function of fluence (black). 
We used the {\sl Chandra} X-ray Center tool \texttt{colden} to look up the $\NH$ value from HI surveys, in order to estimate the optical depth of X-ray scattering at 1 keV. We then multiplied the fluence of each flare by a factor of $(1 - e^{-\tau})$ to estimate the integrated fluence of the resulting dust echo (orange).

A higher instrument background makes it difficult to observe a dust scattering halo. A small fluence (relative to the quiescent state) will produce a small perturbation in the scattering halo brightness that is unlikely to be observable. To avoid modeling the problem, we use the examples of spectacular dust ring echoes from the literature (bottom portion of Table~\ref{tab:flares}) to arrive at approximate thresholds for observation with modern-day X-ray telescopes. 
We chose flares with $\log \fcgs > -2$ as candidates for producing dust echo rings. However, some of these flares may appear bright due to having a low ISM column. At the same time, bright flares offer the chance to produce serendipitous results in high contrast.  For example, V404~Cygni has $\NH \approx 6 \times 10^{21}$~cm$^{-2}$ or $\tau \approx 0.1$. The corresponding estimate yields the dimmest available dust echo fluence ($\approx 0.006$~erg~cm$^{-2}$) in Table~\ref{tab:flares}, yet V404~Cygni produced some of the clearest multi-structured dust echo rings \citep{Heinz2016}. We take  0.005~erg~cm$^{-2}$ as an approximate threshold for effective dust echo tomography with modern day instruments. 

Because one object can produce multiple flares, we counted the number of MAXI targets that exhibited a flare with fluence larger than a given threshold, yielding the number of sight lines available for dust echo tomography (Figure~\ref{fig:FlareHist}, right). We found that 34 of the objects exhibited flares with $\log \fcgs > -2$, and 24 of these have predicted dust echoes over the 0.005~erg~cm$^{-2}$ threshold during the last 9~years of MAXI operation. 
However, more analysis is needed to determine which of these would have produced the thin, high contrast rings that are ideal for measuring the line-of-sight dust distribution.

\subsection{Identifying Sources of Dust Echo Rings}
\label{sec:rings}

High fluence flares are necessary to produce dust scattering halos, but many of them have a large fluence simply because they are long. Two other conditions are important for identifying dust echo candidates. First, flares must be short enough to produce sharp rings. 
Second, the bursts must be accompanied by a long period of quiescence so that the dust echo rings stand out in contrast to the quiescent dust scattering halo. 
A survey by \citet{Valencic2015} showed that a majority of X-ray scattering halos are dominated by scattering from a single cloud, rather than isotropically distributed dust.
The time delay associated with a particular angle can be inverted to solve for the angle at which a dust scattering echo will appear ($\theta$), from a burst that occurred at some time ($t$) prior to now:
\begin{equation}
	\theta = \left[ \frac{2c~(1-x)~t}{xD}\right]^{1/2}
\end{equation}
where $D$ is the distance to the X-ray source, $x$ is the distance to {\bf a} dust cloud divided by $D$, and $c$ is the speed of light \citep{TS1973}.  This equation can be used to fit dust echo rings with multiple discrete ISM clouds, or, applying a convolution with the flare light curve, can be used to measure contiguous line of sight dust abundances \citep{Heinz2015, Heinz2016}. While clouds or dust material that are extended along the line of sight will alter the perceived thickness and time delay of a dust echo, a full examination of these geometric effects is beyond the scope of this work.

For a fixed dust cloud position, the thickness of a dust echo ring ($\Delta \theta$) will depend on the duration of the flare ($t_f$) so that $\Delta \theta \propto t_f^{1/2}$. In contrast, the dust scattering halo will return to its quiescent state out to some angle, $\theta \propto t_q^{1/2}$ where $t_q$ is the duration of the quiescent period before or following the flare. We do not set a maximum duration for $t_f$. All scattering halos dim in surface brightness at large angular distance from the central source source, so a return to a dim quiescent state will always produce the appearance of rings. This was apparent from the outburst of 4U 1630-47, lasting over 100 days, which produced an 8 arcminute scale ring \citep{Kalemci2018}. Ideal echoes will have thin rings relative to the size of the quiescent halo, requiring $t_f / t_q << 1$. 

\begin{figure*}
\centering
\includegraphics[width=\textwidth]{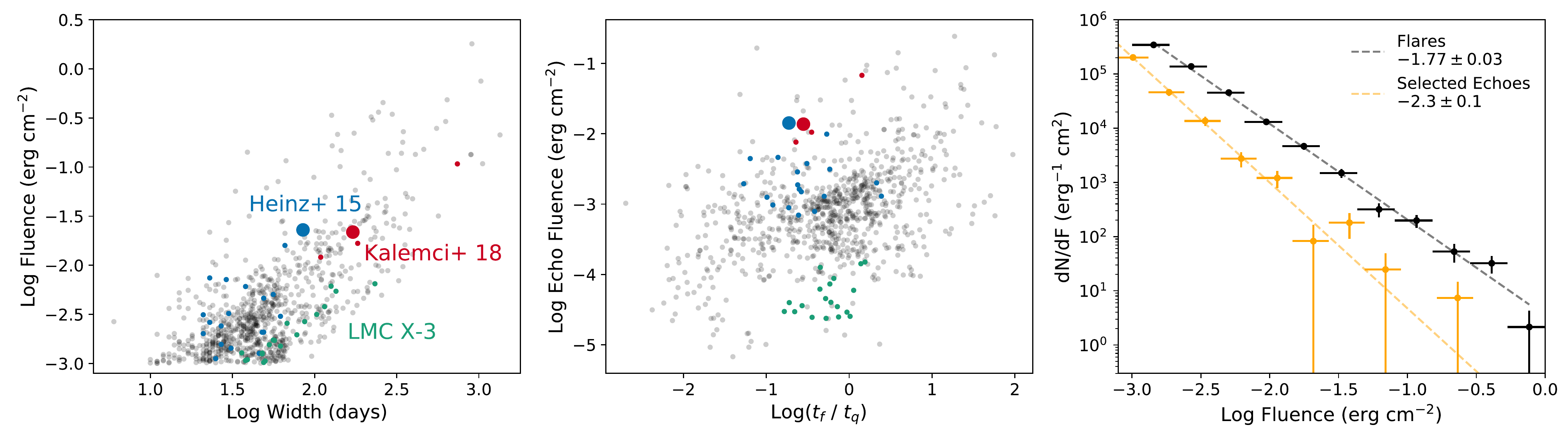}
\caption{
{\sl Left:} Relationship between flare duration and 2-4~keV fluence of the flare, as measured by MAXI. Naturally, the larger fluence flares tend to arise from longer duration flares, up to 1000 days. The blue, red, and green circles show data points from the flares observed from Cir~X-1, 4U~1630-47, and LMC~X-3, respectively. The flares that led to dust echoes studied by \citet{Heinz2015} and \citet{Kalemci2018} are highlighted by the large blue and red circles, respectively.
{\sl Middle:} Plot of the dust echo 2-4~keV fluence versus the ring sharpness metric, $(t_f/t_q)$. The larger fluence flares, which are often longer, are less useful because they fill out more of the dust scattering halo and produce broad rings that are more likely to overlap. The Cir~X-1, 4U~1630-47, and LMC~X-3 flares are highlighted in the same way as the left plot (blue, red, and green). In general, flares with $(t_f/t_q) << 1$ produce the most ideal dust echoes for high resolution dust echo tomography.
{\sl Right:} A power law fit to the fluence distribution for all flares (black) and dust echoes with $t_q/t_f < 1$ (orange) yields power law slopes of $-1.77 \pm 0.03$ and  $-2.3 \pm 0.1$, respectively. 
\label{fig:Interpret}
}
\end{figure*}

Figure~\ref{fig:Interpret} shows the relationship between flare duration, fluence, and $t_f/t_q$ for the flares identified in the MAXI dataset. We determined the duration of the quiescence directly before and after each flare, and chose the larger $t_q$ value. 
Smaller values of the $t_f/t_q$ lead to thinner dust echo rings. The dust echoes arising from the Cir~X-1 flare \citep{Heinz2015} and 4U~1630-47 \citep{Kalemci2018}  
were both high fluence with moderate values of $0.1 < t_f/t_q < 1$. 
The flares from LMC~X-3 are highlighted in Figure~\ref{fig:Interpret} (orange) to demonstrate a population of flares arising from a highly variable source with no persistent quiescent state.
We found nine objects that produced bright $> 10^{-2}$~erg~cm$^{-2}$ flares detected by MAXI with $(t_f/t_q) < 0.1$. Of these, four have an estimated dust echo fluence $> 0.005$~erg~cm$^{-2}$. These four have not been followed up or published: LS~I~+61~303, V*~BQ~Cam, XTE~J1752-223, and MAXI~J1535-571.

Figure~\ref{fig:Interpret} (right) shows the number density of flares as a function of 2-4~keV fluence, which follows a power law of slope $-1.77 \pm 0.03$ (black).  We also calculated the fluence distribution for dust echoes with $t_f/t_q < 1$ (orange). The distribution follows a power law for $\log \fcgs > -3.0$ and is flat for lower fluences, due to our sensitivity limit. 
For this reason, we limit analysis to the 
$\log \fcgs > -3.0$ dust echo distribution, which fits with a power law of slope $-2.3 \pm 0.1$. 
In the next section, we use this trend to estimate the number of dust echoes that will be seen by the next generation of X-ray observatories.

\subsection{Avenues for Future Study}
\label{sec:future}

\begin{figure}
\centering
\includegraphics[scale=0.75]{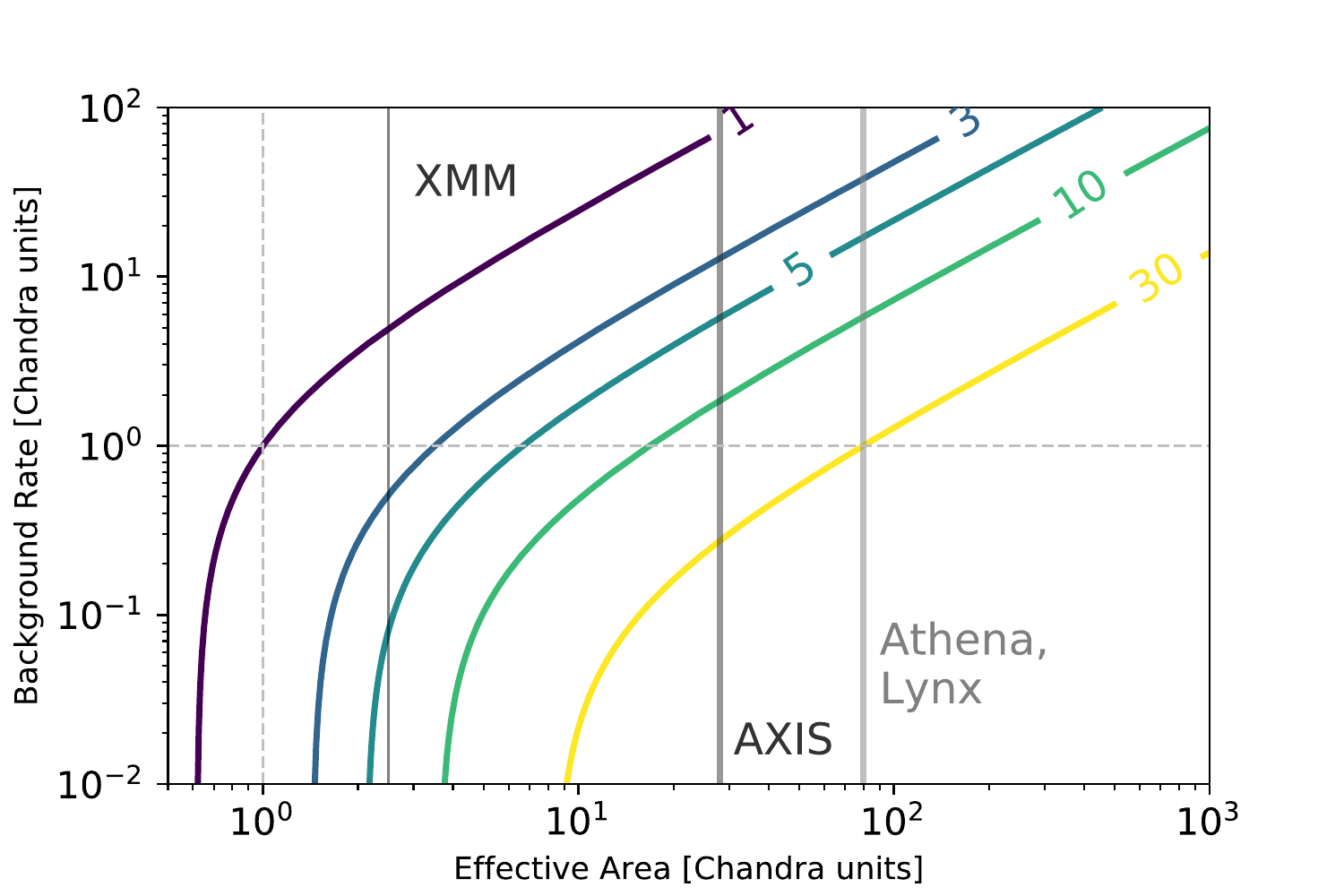}
\caption{
Predicted number of dust scattering echoes observable by different telescopes, depending on the effective area and background levels, as compared to \Chandra. For the next generation of telescopes ({\sl Athena}, {\sl Lynx}, and AXIS) we expect about 30 times more dust echoes than observable with current instruments, depending on the background levels achieved.
\label{fig:SensitivityCurves}
}
\end{figure}

In the future, more sensitive X-ray telescopes will extend dust echo tomography to dimmer flares, opening up more sight lines for probing the 3D distribution of dust via X-ray scattering. 
For a fixed exposure time, we solved for the fluence ($f$) for which the signal-to-noise ratio is the same as the signal-to-noise ratio for a telescope with no background ($f_0$).
\begin{equation}
\label{eq:fluxthreshold}
	f = \frac{f_0}{2} \times \left(1 + \sqrt{1 + 4b/f_0}\right)
\end{equation}
where $b$ is the background surface brightness.\footnote{The background is due to a combination of instrumental, charged particle, and cosmic X-ray background, which change with time and position on the sky. \Chandra\ has relatively low,  stable, and well documented background rates compared to other currently active X-ray telescopes, making \Chandra\ a good baseline for comparison.} 
Since any flux threshold is inversely proportional to the effective area ($a$), we substitute $f_0$ with $1/a$ in Equation~\ref{eq:fluxthreshold}. We calculated $f$ for a grid of effective areas and backgrounds, relative to \Chandra.  We then calculated the total number of scattering ring echoes expected ($N$) by integrating the predicted fluence density distribution for echoes, $dN/df$ (Figure~\ref{fig:Interpret}, orange), extrapolating the power law to fluences with $\log \fcgs < -3$. 
Figure 5 shows several contours for $N$, predicting the observable number of high signal-to-noise scattering ring echoes, relative to the {\sl Chandra} effective area and background surface brightness.

{\sl Athena}, expected to launch around 2030, will have a 1~keV effective area of 2~m$^2$ \citep{Athena2017}, approximately 80 times the current soft X-ray effective area for {\sl Chandra} ACIS-I. 
{\sl Athena} will thereby observe on the order of 30 times more dust echoes than {\sl Chandra} can, depending on the instrument background levels. The concept mission, {\sl Lynx}, will have a similar effective area to {\sl Athena} with the imaging resolution of {\Chandra}. The Advanced X-ray Imaging Satellite (AXIS) concept mission has a proposed  1~keV effective area of 7000~cm$^{2}$ with 10-20 times lower background than \Chandra\ \citep{AXISspie}. Thus AXIS would be able to image a similar number of dust echoes to {\sl Athena} and {\sl Lynx}.

The increased sensitivity offered by the next generation of X-ray telescopes will also constrain the abundance and distribution of dust in the intergalactic medium (IGM) through dust scattering echoes left behind by previously active galactic nuclei (AGN). Using the formulations of \citet{Corrales2015a},  a telescope with ten times the {\sl Chandra} sensitivity will be able to image IGM scattering echoes on the order of $20''$ - $80''$ in radius, corresponding to AGN activity $\sim 10^2$ - $10^3$~years prior. Using the numbers of bright $z > 1$ AGN visible from all-sky surveys, the number of echoes one can expect to find in the entire sky is:
\begin{equation}
	N^{\rm IGM}_{\rm ech} \sim 10-100 
    	\left( \frac{\nu_{fb}}{10^{-3}~{\rm yr}^{-1}} \right)
\end{equation}
where $\nu_{fb}$ is the characteristic frequency for rapid quenching of an AGN accretion flow. We refer the reader to the original work of \citet{Corrales2015a} for a detailed discussion on how AGN variability and feedback can be constrained by IGM dust echoes.

\section{Conclusions} \label{sec:conclusions}

Examination of nine years of MAXI light curves reveals 34 objects that exhibited bright X-ray flares with fluences $> 10^{-2}$~erg~cm$^{-2}$, with durations $\sim 30$-300~days. By comparing the flare duration to the time in quiescence, we estimate that nine of these were short enough to produce sharp ring echoes: approximately one candidate per year.  Using $\NH$ to estimate the dust echo brightness, four of the flares might have produced dust echo rings detectable by current X-ray telescopes.  Only one of these sight lines, Cir~X-1, has been imaged and studied in detail.

With the next generation of X-ray telescopes, dust ring echoes will become common features of the Galactic ISM.
We expect {\sl Athena}, {\sl Lynx}, and AXIS to be $\geq 30$ times more sensitive to dust echoes in comparison to \Chandra. The result will be hundreds of time-variable X-ray scattering halos. 
Of these, we expect $\sim$10-30 sharp dust ring echoes per year, which are ideal for determining 3D distributions of ISM dust with the detail of \citet{Heinz2015,Heinz2016}.

Despite this work focusing on thin ring dust echoes, all bright X-ray sources have dust scattering halos that vary with the light curve of the central source. About half of the flares found in this study had $t_f/t_q > 1$. The resulting image will be a blend of broad rings. 
Interpreting these images will require more advanced dust scattering halo timing techniques. The results will open an avenue for mapping Galactic and intergalactic structures in an entirely new way.

\acknowledgments
We wish to thank the anonymous referee for their thoughtful comments that greatly improved the clarity of the paper. 
This research has made use of MAXI data provided by RIKEN, JAXA, and technical support from the MAXI team.  Support for this work was provided by NASA through Einstein Postdoctoral Fellowship grant number PF6-170149 awarded by the Chandra X-ray Center (CXC), which is operated by the Smithsonian Astrophysical Observatory for NASA under contract NAS8-03060. Additional support for this work came from CXC through grant number TM6-17010X.

\vspace{5mm}
\facilities{MAXI \citep{MAXI2009}}

\software{astropy \citep{Astropy2}, 
		  SciPy, 
          NumPy
          }

\bibliographystyle{aasjournal}
\bibliography{references_new}

\end{document}